\documentclass[]{tEWA2e}
\usepackage{color}
\usepackage[mathscr]{euscript}

\begin{document}

\doi{10.1080/0950034YYxxxxxxxx}



\title{Dispersion peculiarities of hybrid modes in a circular waveguide filled by a composite gyroelectromagnetic medium}

\author{Vladimir R. Tuz$^{1,2,3}$, Illia V. Fedorin$^{4,\ast}$\thanks{$^\ast$Corresponding author. Email: fedorin.ilya@gmail.com \vspace{6pt}}, Volodymyr I. Fesenko$^{3}$, Hong-Bo Sun$^{2}$, \\ Valeriy M. Shulga$^{1,3,5}$, and  Wei Han$^{1,5}$ \\\vspace{6pt}
$^{1}${\em{International Center of Future Science, Jilin University, 2699  Qianjin Street, Changchun 130012, People's Republic of China}}; $^2${\em{State Key Laboratory on Integrated Optoelectronics, College of Electronic Science and Engineering, Jilin University, 2699 Qianjin Street, Changchun 130012, People's Republic of China}}; $^{3}${\em{Institute of Radio Astronomy of National Academy of Sciences of Ukraine, 4, Mystetstv St., Kharkiv 61002, Ukraine}}; $^{4}${\em{National Technical University `Kharkiv Polytechnical Institute', 21, Kirpichova Street, Kharkiv 61002, Ukraine}};  $^{5}${\em{College of Physics, Jilin University, 2699  Qianjin Street, Changchun 130012, People's Republic of China}} \\\vspace{6pt}
\received{November 2016} }

\maketitle

\begin{abstract}
Extraordinary dispersion features of a circular waveguide filled by a longitudinally magnetized composite gyroelectromagnetic medium are studied. The composite medium is considered to be constructed by juxtaposition together of magnetic and semiconductor layers providing all characteristic dimensions of the resulting multilayered system are much smaller than the wavelength in the corresponding part of the composite medium. The waveguide dispersion equation and its eigenmodes are derived. The mode classification is made in a standard manner so that the guided modes are sorted into a class of hybrid HE and EH waves in terms of their unique dispersion characteristics. The numerical results are obtained in the band near the frequencies of corresponding resonances in  constitutive materials of the composite medium, and they show that the modes behaviors  become to be quite diverse due to manifestation of a strong combined geometrical and material dispersion related to the waveguide parameters and gyroelectromagnetic filling, respectively.
\bigskip

\bigskip
\vspace{12pt}

\end{abstract}

\section{Introduction}

The Faraday rotation systems that are based on circular (uniaxial)  waveguides partially or completely filled by a \textit{gyromagnetic} medium (e.g. a longitudinally magnetized ferrite) are important amplitude and phase control devices in microwaves through decades. To date, the methods of characterization of such gyrotropic systems are well established \cite{gamo_JPSJ_1953, kales_JApplPhys_1953, Waldron_RadioEng_1959, donovan_BJApplPhys_1967, Che_JEMWA_2002, Cojocaru_JOptSocAmB_2010} and have been summarized in a set of subsequent textbooks \cite{Gurevich_book_1963, thourel_book_1964, fuller_book_1987, Collin_book_1991}. In theory, these methods imply  solution of a boundary value problem providing derivation and classification of the waveguide's (eigen)modes with corresponding identification of their cutoff frequencies and dispersion peculiarities. Remarkably, for particular modes of the gyrotropic waveguide, alongside with the geometrical dispersion, the material dispersion related to the magnetic material function (e.g. permeability) can appear to be a dominant feature which is associated with manifestation of the ferromagnetic resonance \cite{Tuz_JOpt_2010}.

Apart waveguides with gyromagnetic filling, similar systems with \textit{gyroelectric} filling are also well known \cite{Kuno_MTT_1967, liu_PIER_2000}, among which plasma-filled circular waveguides should be mentioned \cite{Maraghechi_PhysPlasmas_1994, Hwang_PhysPlasmas_1998}. The material dispersion of such gyroelectric waveguides are influenced by manifestation of cyclotron and plasma resonances which are  related to the dielectric material function (e.g. permittivity or conductivity). Interest in studying plasma-filled waveguides has arisen due to progress in high-frequency electronics with the objective to create powerful microwave amplifiers and generators assuming their application to high-resolution and imaging radars, and systems for particle accelerators and fusion.

Nowadays, the main research efforts in the field of circular gyrotropic waveguides are aimed at utilization of modern artificial gyrotropic composites as the waveguide filling \cite{Xu_JEMWA_2009, Baqir_Pier_2011, Dong_Pier_2012} including those suitable for transmission in the range of  terahertz waves \cite{Novotny_PhysRevE_1994, Fesenko_2005, Baida_PhysRevB_2006, Atakaramians_AdvOptPhoton_2013}. The theory of such artificial materials involves simultaneous presence of both gyromagnetic and gyroelectric effects \cite{prati_JEMWA_2003, Tuz_PIERB_2012, Tuz_JO_2015, Tuz_Springer_2016, Tuz_JMMM_2016, Fesenko_OptLett_2016, Baibak_2014} in order to reach a wide diversity of electromagnetic features of waveguide systems. It should be noted, in the subject field of interest, in recent years such \textit{gyroelectromagnetic} composites are usually considered within the conception of metamaterials, in the framework of which they are widely discussed from the viewpoint of achieving below-cutoff and backward waves propagation in miniaturized circular waveguides \cite{Novitsky_JOpt_2005, Brand_IntJElectron_2009, Ghosh_Electromagn_2012, Pollock_MTT_2016, Pollock_AppPhys_2016}. 

In order to realize such a gyroelectromagnetic filling of actual microwave waveguide systems, it is proposed to use a low temperature magnetized plasma admixed by micron ferrite grains (e.g. yttrium iron garnet) \cite{Rapoport_PhysPlasmas_2010}. In this mixture the material dispersion of permeability is caused by the high frequency magnetization of the grain subsystem which is important in the vicinity of a ferromagnetic resonance. The latter is chosen to be located near the electron cyclotron frequency of plasma that is responsible for the material dispersion of permittivity. Since the resulting medium is under an action of the static magnetic field, the combined gyroelectromagnetic effect arises.

As an alternative, gyroelectromagnetic behaviors can be achieved by combining  some magnetic and semiconductor materials into a uniform composite structure constructing, for example, a superlattice \cite{Datta_SuperMicro_1985} or assembling components at the molecular or atomic level to obtain some polycrystalline forms \cite{Akyurtlu_TeraSci_2013, dong_book_2016}. Remarkably, it is reported \cite{Jungwirth_RevModPhys_2006} that such magnetic-semiconductor heterostructures can exhibit a combined gyroelectromagnetic effect from gigahertz up to tens of terahertz.

For the waves propagating in any kind of an unbounded gyrotropic medium being in the Faraday configuration (i.e. it can be a gyroelectric medium described by permittivity tensor, a gyromagnetic medium described by permeability tensor as well as a gyroelectromagnetic medium described by both permittivity and permeability tensors) there are two distinct circularly polarized modes that have only transverse field components, whereas in any circular gyrotropic waveguides they cannot be either pure transverse electric (TE) or pure transverse magnetic (TM) ones. For a mode to exist in a circular gyrotropic waveguide, the longitudinal components of both the electric and magnetic fields must be coupled, therefore, in general, the electromagnetic field has all six components. Such waves are classified as \textit{hybrid} EH-modes and HE-modes \cite{fuller_book_1987}, and these modes emerge as some superposition of the longitudinal and transverse waves. 

Since waves of a circular gyrotropic waveguide have a hybrid character, it is obvious, that both mode composition and electromagnetic field pattern appear to be rather complicated. The complexity also increases significantly in the frequency bands where the material functions of the waveguide's filling manifest their resonant nature that leads to appearing a considerable combined geometrical and material dispersion. Therefore, the goal of this paper is to highlight the main dispersion peculiarities of the hybrid waves in the circular waveguide which is completely filled by a gyroelectromagnetic composite medium, namely, to classify the waveguide modes and identify regions of their normal as well as anomalous dispersion within the band near the frequencies of ferromagnetic and plasma resonances.

\section{Outline of problem}
\label{sec:problem}

\begin{figure}[htbp]
\centering
\includegraphics[width=0.6\linewidth]{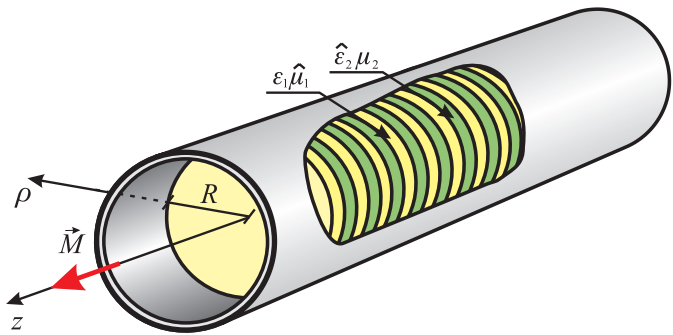}
\caption{Schematic of a completely filled circular waveguide bounded by a perfectly conducting wall. The filling is a composite medium comprising periodically alternating magnetic and semiconductor layers. The whole waveguide system is influenced by an external static magnetic field applied in the Faraday geometry. (The colour version of this figure is included in the online version of the journal.)}
\label{fig:fig_1}
\end{figure}

Therefore, further we consider a cylindrical waveguide of radius $R$ which is completely filled by a composite medium (Fig.~\ref{fig:fig_1}). This composite medium is constructed by juxtaposition together of magnetic and semiconductor layers that are arranged periodically along the $z$-axis.  It is supposed that all characteristic dimensions (i.e. layers thicknesses and the period length) of such multilayered structure are much smaller than the wavelength in the corresponding part of the composite (the long-wavelength limit). The resulting waveguide structure is influenced by an external static magnetic field $\vec M$ which is aligned along the axis of the guide, i.e. along the $z$ direction (the Faraday geometry). We consider that the strength of this magnetic field is high enough to form a homogeneous saturated state of magnetic as well as semiconductor subsystems.

With taking into consideration the long-wavelength limit in the composite constituents and involving averaging (homogenization) procedures from the  effective medium theory (is omitted here; for details, see, \cite{Agranovich_SolidStateCommun_1991, Tuz_JMMM_2016}), the filling composite medium of the waveguide structure under study is further described with two tensors of relative effective permittivity $\hat\varepsilon$ and relative effective permeability $\hat\mu$ written in the form:
\begin{equation}
 \hat \varepsilon=\left( {\begin{matrix}
   {\varepsilon} & {i\varepsilon_a} & 0 \cr
   {-i\varepsilon_a} & {\varepsilon} & 0 \cr
   0 & 0 & {\varepsilon_\|} \cr
\end{matrix}
} \right),~~~~~
\hat\mu=\left( {\begin{matrix}
   {\mu} & {i\mu_a} & 0  \cr
   {-i\mu_a } & {\mu} & 0  \cr
   0 & 0 & {\mu_\|}  \cr
 \end{matrix}
} \right), \label{eq:eff}
\end{equation}
where the complete expressions for the tensors components derived via underlying constitutive parameters of magnetic and semiconductor layers one can find in \cite{Tuz_JMMM_2016}. In particular, these parameters are calculated in the microwave part of spectrum for a magnetic-semiconductor structure made in the form of a barium-cobalt/doped-silicon superlattice \cite{Wu_JPhysCondensMatter_2007}. Remarkably, for such a structure the characteristic resonant frequencies of underlying constitutive magnetic and semiconductor materials appear to be different but rather closely spaced within the same frequency band. 

Therefore, the problem is reduced to the characterization of the circular waveguide filled by an anisotropic (gyroelectromagnetic) uniform medium whose optical axis is directed in the longitudinal direction of the guide. The method of solution of this problem comprises obtaining the wave equations with respect to the longitudinal components of the electromagnetic field inside the gyrotropic waveguide, and then involving the boundary conditions on the waveguide's perfectly conducting wall. The derivation of this method has been discussed in a set of papers and textbooks \cite{gamo_JPSJ_1953, kales_JApplPhys_1953, Waldron_RadioEng_1959, donovan_BJApplPhys_1967, Che_JEMWA_2002, Cojocaru_JOptSocAmB_2010, Gurevich_book_1963, thourel_book_1964, fuller_book_1987, Collin_book_1991}, therefore, we provide only a brief outline here, based on the treatment given in \cite{Gurevich_book_1963}. 

Starting from Maxwell's equations, in cylindrical polar coordinates $(\rho,\varphi,z)$, a pair of coupled Helmholtz wave equations for the longitudinal field components is written in the form:
\begin{equation}
\begin{split}
&\left(\nabla^2_\bot + a_e \frac{\partial^2}{\partial^2 z} + b_e\right)\mathscr{E}_z+c_e\frac{\partial}{\partial z}\mathscr{H}_z = 0, \\
&\left(\nabla^2_\bot + a_h\frac{\partial^2}{\partial^2 z} + b_h\right)\mathscr{H}_z - c_h\frac{\partial}{\partial z}\mathscr{E}_z = 0,
\end{split}
\label{eq:eqlongitud}
\end{equation}
whose coefficients are: 
$$a_e=\frac{\varepsilon_\|}{\varepsilon},~~b_e=k_0^2\varepsilon_\|\mu_\bot,~~c_e=k_0\mu_\|\left( \frac{\varepsilon_a}{\varepsilon}+\frac{\mu_a}{\mu}\right),$$
$$a_h=\frac{\mu_\|}{\mu},~~b_h=k_0^2\mu_\|\varepsilon_\bot,~~c_h=k_0\varepsilon_\|\left( \frac{\varepsilon_a}{\varepsilon}+\frac{\mu_a}{\mu}\right),$$
where $\mu_\bot = \mu-\mu_a^2/\mu$ and $\varepsilon_\bot = \varepsilon-\varepsilon_a^2/\varepsilon$ are introduced as effective transverse permeability and effective transverse permittivity, respectively, and the operator $\nabla^2_\bot$ is related to transverse components: $$\nabla^2_\bot\equiv\frac{\partial^2}{\partial \rho^2}+\frac{1}{\rho}\frac{\partial}{\partial \rho }+\frac{1}{\rho^2}\frac{\partial}{\partial \varphi}.$$
Further set of equations~(\ref{eq:eqlongitud}) is rearranged into a single fourth-order equation via the next substitutions \cite{Gurevich_book_1963}:
\begin{equation}
\mathscr{E}_z=i\frac{\partial}{\partial z} W\Psi,~~~~
\mathscr{H}_z=i P\Psi,
\label{eq:substitution}
\end{equation}
yielding it in the operator form:
\begin{equation}
\mathscr{L}(\Psi)=0,
\label{eq:waveequations1}
\end{equation}
where $W$ and $P$ are given in Appendix \ref{sec:app_a}, and the operator $\mathscr{L}$ is
\begin{equation}
\begin{split}
\mathscr{L}  \equiv \nabla^4_\bot + \frac{\varepsilon_\|\mu_\|}{\varepsilon\mu}\frac{\partial^4}{\partial z^4} & + \left(\frac{\varepsilon_a}{\varepsilon} + \frac{\mu_a}{\mu}\right)\nabla^2_\bot\frac{\partial^2}{\partial z^2} + k_0\left(\varepsilon_\|\mu_\bot+\mu_\|\varepsilon_\bot\right)\nabla^2_\bot \\
& + 2k_0^2\varepsilon_\|\mu_\|\left(1+\frac{\varepsilon_a\mu_a}{\varepsilon\mu}\right)\frac{\partial^2}{\partial z^2} + k_0^4\varepsilon_\|\mu_\|\varepsilon_\bot\mu_\bot.
\end{split}
\label{eq:operator1}
\end{equation}
If we assume that all field components vary harmonically as $\exp(i\gamma z)$, where $\gamma$ is the propagation constant, the field function $\Psi$ can be represented as a product of two independent transverse and longitudinal functions
\begin{equation}
\Psi = \psi(\rho,\varphi)\mathscr{Z}(z),
\label{eq:twofunctions}
\end{equation}
and differential operator $\partial^2/\partial z^2$ in (\ref{eq:operator1}) should be substituted by the coefficient $(-\gamma^2)$. These lead to representation of the operator $\mathscr{L}$ in the form:
\begin{equation}
\mathscr{L} \equiv \nabla^4_\bot + p\nabla^2_\bot + q,
\label{eq:operator2}
\end{equation}
where 
\begin{equation}
\begin{split}
p &= k_0^2\left(\varepsilon_\|\mu_\bot+\mu_\|\varepsilon_\bot\right)-\gamma^2\left(\frac{\varepsilon_\|}{\varepsilon} + \frac{\mu_\|}{\mu}\right), \\
q &= \varepsilon_\|\mu_\|\left[k_0^4\varepsilon_\bot\mu_\bot-2k_0^2\gamma^2\left(1+\frac{\varepsilon_a\mu_a}{\varepsilon\mu}\right)+\frac{\gamma^4}{\varepsilon\mu}\right].
\end{split}
\notag
\end{equation}
In turn, operator (\ref{eq:operator2}) can be substituted as the product of two terms:
\begin{equation}
\mathscr{L} \equiv \left(\nabla^2_\bot+\mbox{\ae}_{1}^2\right)\left(\nabla^2_\bot+\mbox{\ae}_{2}^2\right),
\label{eq:operator3}
\end{equation}
where numbers $\mbox{\ae}_{1}^2$ and $\mbox{\ae}_{2}^2$ satisfy the following set of equations:
\begin{equation}
\mbox{\ae}_{1}^2 + \mbox{\ae}_{2}^2 = p,~~~~
\mbox{\ae}_{1}^2 \mbox{\ae}_{2}^2 = q.
\label{eq:kappasystem}
\end{equation}
This system is related to biquadratic equation 
\begin{equation}
\mbox{\ae}^4 - p\mbox{\ae}^2 + q = 0,
\end{equation}
whose solution is trivial:
\begin{equation}
\begin{split}
\mbox{\ae}_{1,2}^2&=\frac{1}{2}\left[k_0^2\left(\varepsilon_\|\mu_\bot + \mu_\|\varepsilon_\bot\right) - \gamma^2\left(\frac{\varepsilon_a}{\varepsilon} + \frac{\mu_a}{\mu}\right)\right] \\
&\pm\left\{\frac{1}{4}\left[k_0^2\left(\varepsilon_\|\mu_\bot - \mu_\|\varepsilon_\bot\right) - \gamma^2\left(\frac{\varepsilon_a}{\varepsilon} - \frac{\mu_a}{\mu}\right)\right]^2 +  \gamma^2k_0^2\varepsilon_\|\mu_\|\left(\frac{\varepsilon_a}{\varepsilon} + \frac{\mu_a}{\mu}\right)^2\right\}^\frac{1}{2}.
\end{split}
\label{eq:wavenumbers}
\end{equation}

Applying the operator $\mathscr{L}$ (which is now independent on $z$) to the function $\psi$, we arrive at a set of two wave equations
\begin{equation}
\nabla^2_\bot \psi + \mbox{\ae}_{1,2}^2\psi = 0,
\label{eq:waveequations2}
\end{equation}
which allows us to derive a solution for the waveguide system in the form:
\begin{equation}
\Psi=\sum_{k=1,2} \mathscr{A}_k\mathscr{I}_n(\mbox{\ae}_{k},\rho) \exp\left[i(\pm n\varphi-\gamma z)\right], \label{eq:solution}
\end{equation}
where $\mathscr{I}_n(.)$ denotes the Bessel function of the first kind of order $n$ ($n=0,1,2,...$), upper sign `$+$' is related to the right-handed circularly polarized waves, whereas the lower sign `$-$' is related to the left-handed circularly polarized waves, and $\mathscr{A}_1$ and $\mathscr{A}_2$ are constants to be defined from the boundary conditions  applied to the tangential field components $\mathscr{E}_z$, $\mathscr{E}_\varphi$, $\mathscr{H}_z$, and $\mathscr{H}_\varphi$. The complete expressions of all field components are given in  Appendix~\ref{sec:app_b}.

Imposing the boundary conditions on the waveguide's perfectly conducting walls $\mathscr{E}_z=0$ and $\mathscr{E}_\varphi=0$ at $\rho = R$ we arrive at a set of homogeneous linear equations with respect to coefficients $\mathscr{A}_1$ and $\mathscr{A}_2$:
\begin{equation}
\begin{split}
\sum_{k=1,2} \mathscr{A}_k W_k\mathscr{I}_n(\mbox{\ae}_{k},R) &= 0, \\
\sum_{k=1,2} \mathscr{A}_k\left[T_k\mbox{\ae}_{k}\mathscr{I}'_n(\mbox{\ae}_{k},R) + S_k\frac{m}{R}\mathscr{I}_n(\mbox{\ae}_{k},R)\right] &= 0,
\end{split}
\label{eq:besselsystem}
\end{equation}
where coefficients $T_k$ and $S_k$ are given in Appendix~\ref{sec:app_a}. The vanishing determinant of (\ref{eq:besselsystem}) gives us the dispersion equation with respect to the propagation constant $\gamma$ which determines the eigenmodes of the circular waveguide filled by a gyroelectromagnetic medium under study.

\section{Modes classifications and dispersion features \label{sec:results}}

\begin{figure}[htbp]
\centering
\includegraphics[width=\linewidth]{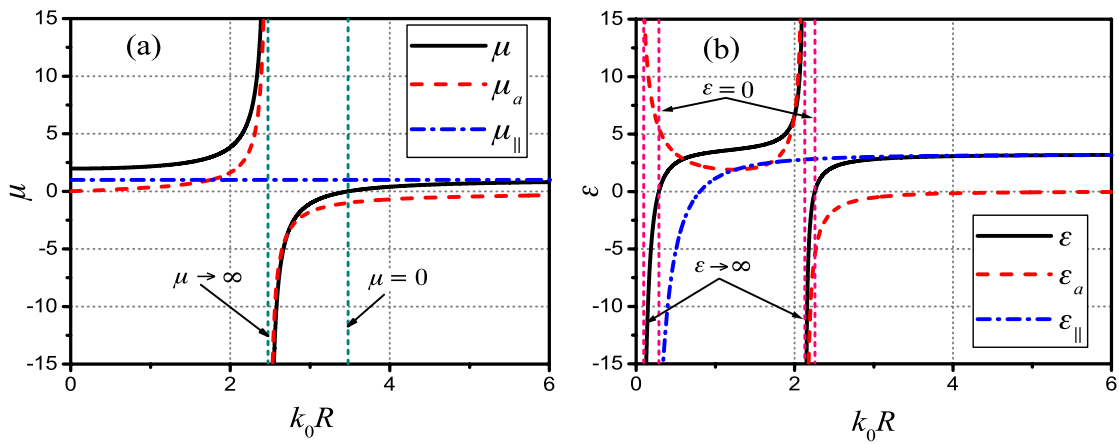}
\caption{Dispersion curves of the tensors components of (a) relative effective permeability $\hat\mu$ and (b) relative effective permittivity $\hat\varepsilon$ of the filling composite medium. For the magnetic constitutive layers, under saturation magnetization of 2000~G, parameters are: $f_0=\omega_0/2\pi=4.2$~GHz, $f_m=\omega_m/2\pi=8.2$~GHz, $\varepsilon_m=5.5$. For the semiconductor constitutive layers, parameters are: $f_p=\omega_p/2\pi=4.9$~GHz, $f_c=\omega_c/2\pi=4.7$~GHz, $\varepsilon_l=1.0$, $\mu_s=1.0$. The ratio of the layers thicknesses is: $d_m/L=d_s/L=0.5$. (The colour version of this figure is included in the online version of the journal.)}
\label{fig:fig_2}
\end{figure}

It is obvious that dispersion features of any waveguide system depend on both geometrical parameters of the guide and electromagnetic properties of its filling material. Especially the latter are important, if the material functions of the underlying constituents of the filling material possess some resonant characteristics that can result in appearance of strong material dispersion of the guided modes. Therefore, in order to clarify the combined effect of geometrical and material parameters on  dispersion features of the structure under study, the frequency band of interest and parameters of the waveguide system are chosen in such a manner as to include characteristic resonant frequencies of both magnetic and semiconductor subsystems of the gyroelectromagnetic composite medium. Moreover, the resonant frequencies of both subsystems are considered to be different, but nevertheless closely spaced within the same frequency band. For further reference, the corresponding dispersion curves of the tensors components of relative effective permeability $\hat\mu$ and relative effective permittivity $\hat\varepsilon$ of the filling composite medium are presented in Fig.~\ref{fig:fig_2} on the frequency scale normalized on the waveguide radius. Two particular frequencies are important for our study, namely, those where the magnetic and dielectric functions reach zero and infinity values. Exactly these extreme states are highlighted in Fig.~\ref{fig:fig_2} with vertical short-dashed lines. We also should note, since we are only interested in the eigenwaves propagation, the effects of losses in the underlying materials of the filling composite medium are excluded from the consideration.

\begin{figure}[htbp]
\centering
\includegraphics[width=\linewidth]{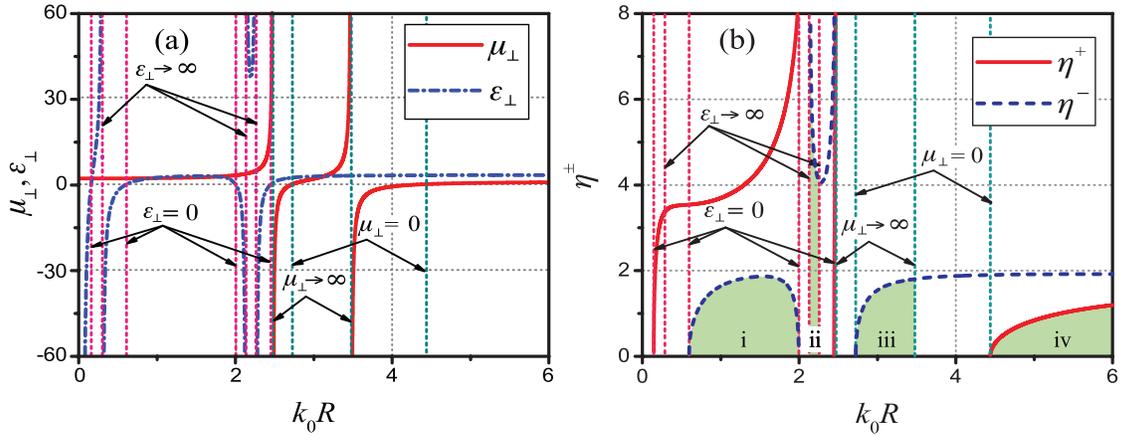}
\caption{Dispersion curves of (a) effective transverse permeability $\mu_\bot$ and effective transverse permittivity $\varepsilon_\bot$, and (b)  the real part of effective refractive index $\eta^\pm$ related to the right-handed and left-handed circularly polarized eigenwaves of an unbounded gyroelectromagnetic medium. All constitutive parameters of the magnetic and semiconductor layers are the same as in Fig.~\ref{fig:fig_2}. (The colour version of this figure is included in the online version of the journal.)}
\label{fig:fig_3}
\end{figure}

For further characterization of the waveguide system under study with the objective to explicitly identify the ranges of existence of the waveguide modes, it is appropriate to derive some generalized material functions which inherit properties of the constitutive parameter tensors $\hat\mu$ and $\hat\varepsilon$. These generalized material functions are effective transverse permeability $\mu_\bot$, effective transverse permittivity $\varepsilon_\bot$, and effective refractive index $\eta^\pm=\sqrt{(\mu\pm\mu_a)(\varepsilon\pm\varepsilon_a)}$ of an unbounded gyroelectromagnetic medium. In the latter term the upper sign `$+$' and the lower sign `$-$' are related to the right-handed and left-handed circularly polarized eigenwaves, respectively, and it is well known, that these circularly polarized eigenwaves are propagating ones only if the corresponding refractive index $\eta^\pm$ acquires real numbers. The dispersion curves of the mentioned generalized material functions are depicted in Fig.~\ref{fig:fig_3}, where, as previously, particular extreme states at which the magnetic and dielectric functions $\mu_\bot$ and $\varepsilon_\bot$ reach zero or infinity values are highlighted with vertical short-dashed lines.

In general, considering extreme states of generalized material functions $\mu_\bot$ and $\varepsilon_\bot$ in conjunction with conditions of existence of circularly polarized eigenwaves of the unbounded gyroelectromagnetic medium, the whole frequency band of interest is divided into four specific sub-bands which are outlined in Fig.~\ref{fig:fig_3}b as the areas colored in light green and numbered in Roman numerals. Remarkably, in all these areas both $\mu_\bot$ and $\varepsilon_\bot$ are positive quantities ($\mu_\bot>0$, $\varepsilon_\bot>0$), whereas their underlying constitutive parameters $\mu$ and $\varepsilon$ can possess different signs. Thus, in the corresponding sub-bands the next conditions hold: (i) and (iv) $\mu > 0$ and $\varepsilon > 0$; (ii) $\mu > 0$ and $\varepsilon < 0$; (iii) $\mu < 0$ and $\varepsilon > 0$, and in all these sub-bands the dispersion equation that follows from equation (\ref{eq:besselsystem}) has a set of real solutions. Each of these solutions corresponds to a particular waveguide mode of the  structure under study. Since in the initial scalar wave equations (\ref{eq:eqlongitud}) there is a coupling between $\mathscr{E}_z$ and $\mathscr{H}_z$ field components, the derived waveguide modes should be classified as hybrid ones.

In the hybrid modes classifications two schemes can be distinguished \cite{Veselov_book_1988, whites_report_1989}. According to the first scheme the hybrid mode has either HE-type or EH-type depending on the magnitudes ratio between the longitudinal electric and magnetic fields components. Thus, in the considered structure geometry it is supposed that the mode has HE-type if $\mathscr{H}_z > \mathscr{E}_z$, otherwise, the mode has EH-type if $\mathscr{E}_z > \mathscr{H}_z$. Note, under such a classification the mode type can vary within the same dispersion curve for different values of $\gamma$, and, therefore, this scheme can result in some ambiguity in classification since it depends on how far the mode is from its cutoff.

The second scheme of the hybrid modes classification implies characterization of an auxiliary reference waveguide that is filled by an isotropic homogeneous medium assuming that the modes of such a waveguide are well known and can be defined definitely. More particularly, for the waveguide structure under study it is supposed to start with such an isotropic (non-gyrotropic) case (i.e. considering the limit $M \to 0$) and classifying modes as those of either TE-type or TM-type beginning from their cutoffs, and then gradually increase the gyrotropic parameters of the underlying constituents (i.e. non-diagonal constitutive tensors elements $\mu_a$ and $\varepsilon_a$) of the gyroelectromagnetic medium inside the guide and trace the changing in the propagation constant $\gamma$. In this way, hybrid modes of HE-type and EH-type of the gyroelectromagnetic waveguide appear to be directly associated with corresponding modes of TE-type and TM-type of the auxiliary reference waveguide.

Therefore, based on the above discussed schemes in the present paper we classify the hybrid modes of the gyroelectromagnetic waveguide system under study considering an auxiliary reference waveguide that has a similar isotropic filling (i.e. for the filling of this reference waveguide we guess that both $\mu_a$ and $\varepsilon_a$ are close to zero), assuming that near the cutoff frequency TE and TM modes of the reference waveguide turn into HE and EH modes of the gyroelectromagnetic waveguide system, respectively. 

\begin{figure}[htbp]
\centering
\includegraphics[width=\linewidth]{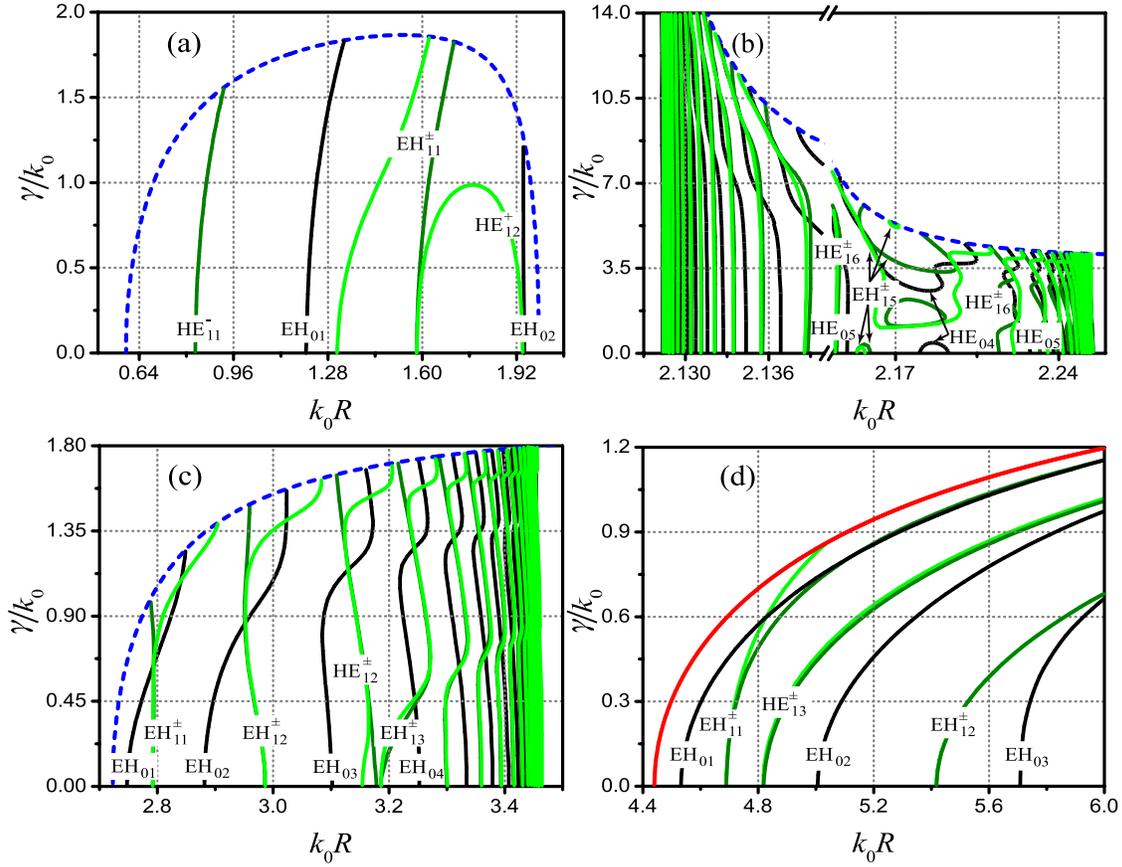}
\caption{Dispersion curves of the lowest types ($n=0,\pm 1$) of hybrid HE$_{nm}^\pm$ and EH$_{nm}^\pm$ modes of a circular waveguide completely filled by a gyroelectromagnetic medium. The light green curves and upper sign `$+$' correspond to the modes with right-handed rotation, whereas the dark green curves and upper sign `$-$' correspond to the modes with left-handed rotation.  Panels (a)-(d) correspond to respective sub-bands (i)-(iv) colored in light green in Fig.~\ref{fig:fig_3}b. All constitutive parameters of the magnetic and semiconductor layers are the same as in Fig.~\ref{fig:fig_2}. (The colour version of this figure is included in the online version of the journal.)}
\label{fig:fig_4}
\end{figure}

Furthermore, in the gyroelectromagnetic waveguide system for each field variations in azimuth ($\pm n$) and radius ($m$) there are two independent solutions of the dispersion equation. These solutions constitute corresponding sets of hybrid modes with an asymmetric and symmetric behaviors \cite{Veselov_book_1988}, wherein, among the asymmetric ones, HE$_{nm}^+$ and EH$_{nm}^+$ modes are referred to the waves with right-handed rotation ($n>0$), while HE$_{nm}^-$ and EH$_{nm}^-$ modes are referred to the waves with left-handed rotation ($n<0$), and in the case of symmetric modes ($n=0$) the upper sign `$+$' or `$-$' is absent. Therefore, in accordance with such a hybrid waves taxonomy, the dispersion curves of modes of the lowest types ($n = 0,\pm 1$) are calculated and classified in Fig.~\ref{fig:fig_4}. One can see that the appearance of these dispersion curves differs drastically in outlined sub-bands (i)-(iv). 

Indeed, in sub-bands (i) and (iv) that are located relatively far away from  the frequencies of the ferromagnetic and plasma resonances (see, panels (a) and (d) in Fig.~\ref{fig:fig_4}), the structure under study operates as a convenient circular gyrotropic waveguide \cite{Collin_book_1991}. It means that the dispersion features of both symmetric and asymmetric modes are mainly determined by the geometric parameters of the guide, wherein the cutoff frequencies rises as the index $m$ of modes increases, and the corresponding curves generally exhibit a normal dispersion line. Presence of a gyrotropic filling primarily manifests itself in the fact that the asymmetric modes starting from the same cutoff frequency diverge into two distinct dispersion curves as $\gamma$ increases. This divergence  expresses a difference in propagation behaviors of the waves with right-handed and left-handed rotations influenced by the external static magnetic field, and this effect disappears as the frequency shifts away from the characteristic resonant frequencies of the filling material. 

Another manifestation of the gyrotropic effect is an exchange of critical conditions between two different modes. As generally known \cite{Veselov_book_1988}, such an exchange can appear between neighbor types of waves which have either the same or different number $m$ of the radial field variations (in particular, it can appear between modes HE$_{11}^+$ and EH$_{11}^+$, HE$_{11}^-$ and EH$_{11}^-$, HE$_{11}^-$ and EH$_{12}^-$, etc.), wherein one of these exchanging modes inevitably exhibits an anomalous dispersion feature. Indeed, for the structure under study the exchange of critical conditions between modes EH$_{11}^+$ and HE$_{12}^+$ is found to be in sub-band (i) (see, Fig.~\ref{fig:fig_4}a).

In sub-bands (ii) and (iii) the hybrid waves properties appear to be much more complicated due to the effect of strong material dispersion of the gyroelectromagnetic filling of the waveguide system under study (see, panels (b) and (c) in Fig.~\ref{fig:fig_4}). Indeed, in these sub-bands the dispersion features of hybrid waves are directly related to the properties of the generalized material parameters of the composite structure, namely, effective transverse permittivity $\varepsilon_\bot$ and effective transverse permeability $\mu_\bot$. Moreover, in accordance with the manifestation order on the frequency scale of the characteristic resonances of the filling material constituents, the properties of $\varepsilon_\bot$ mainly influence waveguide dispersion features in sub-band (ii), whereas the properties of $\mu_\bot$ influence those in sub-band (iii).
    
In particular, considering extreme states of $\varepsilon_\bot$ depicted in Fig.~\ref{fig:fig_3}a, one can conclude that sub-band (ii) appears to be outlined on the frequency scale by the lines where $\varepsilon_\bot\to\infty$. Furthermore, in the middle of this sub-band, $\varepsilon_\bot$ possesses a minimum, and from this minimum the value of $\varepsilon_\bot$ gradually increases in opposing sides towards the low-frequency and high-frequency sub-band edges. It results in a significant convergence of the dispersion curves at the sub-band edges with growth of the number $m$ demonstrating the effect of the so-called reverse cutoff points (i.e. the lower order modes occur at higher frequencies \cite{Brand_IntJElectron_2009}) in the direction of the low-frequency edge. Moreover, the curves exhibit a stepwise shape that is formed by sequential branches with normal and anomalous dispersion for both symmetric and asymmetric modes. Remarkable, in the middle of the sub-band these branches may loop back on themselves.

As is already mentioned, the peculiarities of dispersion curves of hybrid modes in sub-band (iii) are mainly determined by characteristics of $\mu_\bot$. Indeed, this sub-band stars from the frequency where $\mu_\bot=0$ and finishes at the extreme state where $\mu_\bot\to\infty$. Since the value of $\mu_\bot$ rises as the frequency increases, the dispersion curves of both symmetric and asymmetric modes converge towards the high-frequency sub-band edge with growth of the number $m$. The curves also exhibit a stepwise shape which stands out against those in other considered sub-bands by gently slopping branches that manifest a flattened dispersion feature. This effect appears due to the polarization redistribution of the power within the cross-section of the waveguide system.

\section{Conclusions\label{sec:conclusions}}

In the present paper a detailed study of the combined geometrical and material dispersion of hybrid modes of a circular waveguide which is completely filled by a longitudinally magnetized gyroelectromagnetic medium is provided. It is revealed that the dispersion features of waves in such a waveguide system differ drastically from those of conventional dielectric, ferrite or plasma filled waveguides. It is shown that simultaneous presence of gyromagnetic and gyroelectric effects in the waveguide system can provide substantial control of the dispersion characteristics and field distributions of the supported modes.

The waves classification is performed according to which the hybrid modes of the waveguide system are separated into a class of HE and EH modes with symmetric and asymmetric behaviors. Extraordinary dispersion features of these modes are revealed and discussed in view of characteristics of the generalized effective constitutive parameters of the unbounded gyroelectromagnetic medium. On this basis, the ranges of both normal and anomalous dispersion of the hybrid waves are clarified as well. 

The discussed features may provide potential applications in the design of novel waveguide systems, such as ultra-thin waveguides and cavity resonators to manipulate across the bands from gigahertz up to terahertz. 

\section*{Disclosure statement}
No potential conflict of interest was reported by the authors.

\bibliographystyle{tEWA}
\bibliography{gyrotropic_waveguide}

\noindent\medskip

\appendices
\section{}
\label{sec:app_a}

The coefficients used here are:
\begin{equation}
\begin{split}
W_k &= -\frac{\varepsilon}{\varepsilon_\|}\left(\frac{\varepsilon_a}{\varepsilon} + \frac{\mu_a}{\mu}\right)\mbox{\ae}^2_{k}, \\
P_k &= T_k\frac{\mbox{\ae}^2_{k}}{k_0\mu_\|}, \\
T_k &= k_0^2\varepsilon\mu_\bot-\gamma^2-\frac{\varepsilon}{\varepsilon_\|}\mbox{\ae}^2_{k}, \\
S_k &= \frac{\varepsilon_a}{\varepsilon}\left(k_0^2\varepsilon\mu_\bot-\frac{\varepsilon}{\varepsilon_\|}\mbox{\ae}^2_{k}\right) + \frac{\mu_a}{\mu}\gamma^2.
\end{split}
\notag
\end{equation}

\section{}
\label{sec:app_b}

The electromagnetic field components are:
\begin{equation}
\begin{split}
\mathscr{E}_\rho &= i\sum_{k=1,2} \mathscr{A}_k \left[S_k\mbox{\ae}_k \mathscr{I}'_n(\mbox{\ae}_{k},\rho) +  T_k\frac{n}{\rho} \mathscr{I}_n(\mbox{\ae}_{k},\rho)\right],\\
\mathscr{E}_\varphi &= -\sum_{k=1,2} \mathscr{A}_k \left[T_k\mbox{\ae}_k \mathscr{I}'_n(\mbox{\ae}_{k},\rho) +  S_k\frac{n}{\rho} \mathscr{I}_n(\mbox{\ae}_{k},\rho)\right],\\
\mathscr{E}_z &= \gamma\sum_{k=1,2}\mathscr{A}_k W_k \mathscr{I}_n(\mbox{\ae}_{k},\rho),\\
\mathscr{H}_\rho &= \gamma\sum_{k=1,2} \mathscr{A}_k \left[M_k\mbox{\ae}_k \mathscr{I}'_n(\mbox{\ae}_{k},\rho) +  N\frac{n}{\rho} \mathscr{I}_n(\mbox{\ae}_{k},\rho)\right],\\
\mathscr{H}_\varphi &= i\gamma\sum_{k=1,2} \mathscr{A}_k \left[N\mbox{\ae}_k \mathscr{I}'_n(\mbox{\ae}_{k},\rho) +  M_k\frac{n}{\rho} \mathscr{I}_n(\mbox{\ae}_{k},\rho)\right],\\
\mathscr{H}_z &= i\sum_{k=1,2}\mathscr{A}_k P_k \mathscr{I}_n(\mbox{\ae}_{k},\rho),
\end{split}
\notag
\end{equation}
where
\begin{equation}
\begin{split}
M_k &= \frac{1}{k_0\mu}\left[k_0^2(\varepsilon\mu + \varepsilon_a\mu_a) - \gamma^2 - \frac{\varepsilon}{\varepsilon_\|}\mbox{\ae}^2_{k}\right], \\
N &= k_0\varepsilon\left(\frac{\varepsilon_a}{\varepsilon} + \frac{\mu_a}{\mu}\right).
\end{split}
\notag
\end{equation}

\label{lastpage}

\end{document}